# Nonaxisymmetric Structures in the Stellar Disks of Galaxies

Hans-Walter Rix[1,2]

*Institute for Advanced Study, Princeton, NJ, 08540*

and

Dennis Zaritsky[1,3]

*Carnegie Observatories, 813 Santa Barbara St., Pasadena, CA, 91101*

January 19, 1995

[1] Hubble Fellow
[2] Current Address: Max-Planck-Institut für Astrophysik,
    Karl-Schwarzschild-Strasse 1, D-85748 Garching bei München
[3] Current Address: Lick Observatory, University of California, Santa Cruz, CA, 95064

**Abstract**

We study the azimuthal structure of the stellar disks of 18 face-on spiral galaxies, using K$'$(2.2$\mu$m)-band photometry to trace the stellar surface mass density. Assuming the disks are co-planar, we characterize their deviation from axisymmetry by the fractional amplitudes, $A_m(R)/A_0(R)$, and phases, $\varphi_m(R)$, of the $m^{th}$ azimuthal Fourier components at radii $R$ about the photometric galaxy center. We find that most disks exhibit a wealth of non-axisymmetric structures, specifically: (1) that about one third of them are substantially lopsided ($A_1/A_0 \gtrsim 0.20$) at 2.5 disk exponential scale length, (2) that almost one half of them have strong two-armed spirals with an arm/interarm surface-brightness contrasts of order unity, and (3) that typical disks have some intrinsic ellipticity. We estimate that in the disk plane the characteristic ellipticity of the underlying potential is $0.045^{+0.03}_{-0.02}$. However, the spiral pattern couples significantly to the estimate of the intrinsic ellipticity, and our measurement may represent an upper limit on the "true" potential triaxiality. We estimate the radial streaming motions of the disk stars, $v_R$, which are produced by these distortions. By averaging over our sample of galaxies and all azimuthal angles, we find $\langle v_R \rangle \sim 7$ km s$^{-1}$ due to lopsided distortions and $\langle v_R \rangle \sim 6$ km s$^{-1}$ due to intrinsic ellipticity. These non-circular motions are expected to contribute $\sim 0.15$mags scatter to measurements of the Tully-Fisher relation.

*Subject Headings:* Galaxies: Kinematics and Dynamics, Photometry, Spiral, Structure – *Infrared*: Galaxies

# 1 Introduction

Stellar disks are, by definition, highly flattened. Most galaxy models also assume that their mass distribution in the disk plane is axisymmetric. However, deviations from axisymmetry may create radial streaming motions and so strongly alter the dynamics and evolution of galaxies. Despite their importance, many basic questions regarding the non-axisymmetric nature of disks remain yet to be answered. Do disks have a significant intrinsic ellipticity[1]? Are spiral arms small ($\ll 1$) or large ($\gtrsim 1$) fractional density enhancements? Is the mass distribution symmetric? Recent advances in image processing techniques and the development of large-format IR array detectors provide an opportunity to re-examine these issues. In this paper, we use these tools to work toward a more complete understanding of galaxy disks by measuring the nonaxisymmetric components of the stellar *mass* distribution for a sample of nearby spiral galaxies.

Much of the previous exploration of the dynamical effects of nonaxisymmetric disks has focused on our Galaxy. Various investigators who attribute the observed HI kinematics within the Milky Way to a *radial* motion of the LSR (Blitz and Spergel 1990; Kuijken 1991) conclude that the Galaxy's disk cannot be stationary and axisymmetric. Kuijken and Tremaine (1994) arrive at a similar conclusion, but without invoking a radial motion for the LSR. Nevertheless, it is difficult to determine the precise form of the required perturbation; the resolution of this problem is complicated by our position within the very disk whose shape is being analyzed.

Realizing that it is often easier to solve one's neighbors' problems than one's own, we turn our attention to the shapes of stellar disks in a sample of nearby galaxies. Several authors (*e.g.* Binney and de Vaucouleurs 1981; Grosbøl 1985) have used the frequency distribution of isophote shapes in order to measure disk ellipticity in external galaxies. These studies rely on the assumption that the sample is unbiased with respect to galaxy inclination (an assumption which is probably violated, e.g. Huizinga and van Albada 1992), and on the measurement of a single optical wavelength isophote shape. The distorting effects of dust and star formation on ellipticity measurements are exacerbated when only one single isophote is used.

---

[1] We use the term "elliptical" throughout, rather than the more precise "triaxial" ($a \neq b$ and $a, b \gg c$). Whenever we use the term elliptical we mean the intrinsic disk ellipticity (within the disk plane) rather than the observed ellipticity arising from projection effects.

Our approach has two key advantages over the previous studies. First, we minimize projection effects on the apparent disk shapes by selecting objects known *a priori* to be nearly face-on on the basis of their kinematics. Second, we examine the surface brightness distribution of these nearby galaxies in the near infrared (K-band, $2.2\mu m$). It is likely that at this wavelength the surface brightness distribution provides a direct measure of the nonaxisymmetric features in the mass distribution. In the grand-design spiral M51 the K-band light is attenuated by $\lesssim 10\%$ even in the most prominent dust lanes, compared to $\sim 50\%$ at $I(0.8\mu m)$, (Rix and Rieke, 1993, hereafter RR93). RR93 also found that despite vigorous star formation in the arms, the luminosity of young supergiants comprises a small fraction of the total K-band light in M51. Because longer wavelength images start being contaminated by emission from hot dust, K-band imaging of face-on galaxies is the most straightforward way of assessing the stellar surface mass density of galaxy disks.

We concentrate on three specific distortions: elliptical disks, lopsided disks, and two-armed spirals. The ellipticity of disks may bear the imprint of details of the galaxy formation process. Simulations suggest that the galaxy formation process is likely to lead to significantly triaxial halos (e.g. Warren *et al.*1992), which may harbor elliptical disks. At radii where the halo dominates the potential, the shape of the disk is straightforwardly related to the shape of the halo potential. Furthermore, the shape of the disk affects the Tully-Fisher (Tully and Fisher 1977) relationship between a galaxy's luminosity and HI linewidth (hereafter referred to as the TF relationship). This relationship is currently a key tool in the study of the local Universe and its small "intrinsic" scatter attests to the homogeneity of the galaxy formation process. Franx and de Zeeuw (1992) demonstrated that even a relatively small amount of disk ellipticity ($\epsilon \lesssim 0.1$) introduces significant scatter into the TF relation and argued in turn that the observed TF scatter provides an upper limit on the ellipticity of the HI disks. The second type of distortion, lopsidedness, may be a clue to recent infall activity or interactions and may affect disk evolution by generating large radial motions. Finally, measuring the mass amplitude of the spiral arms provides a direct indication of the driving force affecting various processes associated with spiral arms, such as gas compression and star formation. The fractional amplitude of the driving mass perturbation is generally assumed to be less than unity (see e.g. Athanassoula 1984, and references therein), even though there is little support for this assumption from the observations (e.g. Schweizer, 1976; Jensen, Talbot and Dufour, 1981; Elmegreen and Elmegreen, 1984, 1985; Elmegreen *et al.*1992, Elmegreen *et al.*1993; Grosbøl, 1993; RR93). However, as of now there is no definitive measurement of the dynamical strength of spiral arms because the presence of dust and star formation in the spiral arms complicate this measurement. A detailed IR surface photometry study of M51 (RR93, see also Schweizer 1976, Elmegreen and Elmegreen, 1984) confirmed that the spiral arm mass perturbation is large (fractional density enhancements of a factor of 2 to 3) in that galaxy. While that result cannot be generalized because of the strong interaction between M51 and NGC 5195, it may indicate that spiral arm mass amplitudes have been significantly underestimated.

This paper is organized as follows. In Section II we discuss the galaxy sample, the observations, and the data reduction. In Section III we describe the azimuthal Fourier decomposition of the galaxy images, present an overview of the empirical results, and describe the model fitting. In Section IV we also discuss the connection between surface brightness and surface mass, the connection between the isophote shapes and the potential distortion, ellipticity of spiral galaxy disks, lopsidedness in galaxies, and the strength of spiral arms. Throughout Section IV we discuss briefly the implications of the observed asymmetries on disk dynamics. Section V presents the conclusions. Appendices provide details on the model fits, the derivation of our estimated distribution of intrinsic disk ellipticities, and a discussion of the relative contribution of young stars to the total light in the near-IR.

# 2 The Data

## 2.1 Sample Selection

The measurement of disk shapes is simplified by obtaining a sample of face-on (zero inclination or $i = 0°$) galaxies. Inclinations are generally determined from a galaxy's apparent axis ratio under the assumption that disks are intrinsically round. To measure intrinsic disk shapes we obviously must use a shape-independent means of estimating the disk orientation. We attempt a purely kinematic determination of the inclination, $i$, by "inverting" the TF relation. In its standard form, the TF relation is given by

$$M_B = M_B^* - \alpha \log\left(\frac{W'_{20}}{2v^* \sin i}\right), \tag{1}$$

with $M_B^* = -19.55$ mag, $\alpha = -7.48$, and $v^* = 158$ km s$^{-1}$ (Pierce and Tully 1992), where $W'_{20}$ represents the width of the HI emission profile at the 20% of the maximum level, corrected for the internal velocity dispersion of the gas. With this relation we assign to each galaxy of absolute magnitude $M_B$ a "characteristic velocity", $v(M_B)$ (which approximates the disk circular velocity). Inverting Eq. (1) to solve for $i$ results in

$$i = \sin^{-1}\left(\frac{W'_{20}}{2v(M_B)}\right). \tag{2}$$

Here we determine a galaxy's inclination by comparing its observed HI line width with its characteristic velocity, derived from its absolute magnitude (This magnitude, in turn, is calculated by assuming pure Hubble flow and $H_0 = 75$ km s$^{-1}$ Mpc$^{-1}$.) Note that for nearly face-on inclinations the observed linewidth is very sensitive to $i$, while the projected shape is not. For example, if a face-on galaxy with $v_{\rm circ} = 200$ km s$^{-1}$ is tilted by 15°, its observable linewidth has already increased to 104 km s$^{-1}$, about a quarter of the true value, while the axis ratio of the projected circular disk has changed only to 0.97. However, even galaxies with perfectly face-on stellar disks may have a linewidth larger than the vertical gas velocity dispersion (Lewis 1984), for example because many gas disks are warped at large radii. As a consequence luminous galaxies with narrow observed HI linewidths are much rarer than expected from a uniform distribution in $\cos i$ and an equation analogous to Eq. 2 (Lewis 1984). We did not restrict ourselves to well known examples or samples of face-on galaxies (e.g. Lewis 1984) because many of these studies are based on a stringent axis-ratio pre-selection of the objects. Instead, we searched the largest databases available, the Huchtmeier-Richter catalog (Huchtmeier and Richter 1989) for HI linewidths and the CfA Redshift database (Huchra 1993) for apparent galaxy magnitudes and redshifts, in which most entries are unrelated to the question at hand (therefore avoiding any morphological bias). For all candidate galaxies with published HI profiles, we examined the original data to assure that they were of good quality and that there was no evidence for a gas-rich companion in the beam. We examined images on POSS or ESO sky survey images to restrict our sample to isolated field spirals with $i \lesssim 25°$.

We list in Table 1 our final sample of 18 galaxies. We present the galaxy's name in column (1), its type in column (2) from the Revised-Shapley Ames catalog (Sandage and Tammann 1981), its apparent blue magnitude and diameter from RC3 (de Vaucouleurs *et al.* 1991) in columns (3) and (4) respectively, its absolute magnitude corresponding to $H_0 = 75$ km s$^{-1}$ Mpc$^{-1}$ in column (5), its recessional velocity and linewidth from the Huchtmeier-Richter HI catalog corrected for an internal velocity dispersion of 10 km s$^{-1}$ (Tully 1988) in columns (6) and (7) respectively, its inclination from Eq. 2 in column (8), and the total K' exposure time in column (9). The median luminosity of galaxies in this sample is $1.1 L^*$ and their median distance is 31 Mpc. Because the catalogs used to identify the sample have heterogeneous selection criteria, our resulting sample is not statistically complete. Nevertheless, there is no obvious correlation between a galaxy's inclusion in the catalogs and the details of its stellar disk shape. In particular, all galaxies with (apparent) axis ratios between 0.85 and 1 should stand a comparable chance of entering the catalogs. Therefore, we do not believe that a bias toward either circular or slightly elliptical disks has been introduced into the sample.

## 2.2 Observations and Data Reduction

We present K$'$(2.2$\mu$m)-band images for all sample galaxies in Figure 1. These are complemented by additional I(0.8$\mu$m) images for six sample galaxies (see Table 1), which were used to construct external error checks. The K$'$-band data were obtained during 1992 Nov 11 through 14 at the Las Campanas 2.5m telescope with a NICMOS3 256×256 array. The K$'$(2.2$\mu$m) filter has an effective wavelength of 2.16$\mu$m and a bandwidth of 0.33$\mu$m and is designed to reduce the background sky flux compared to the standard K filter by lowering the long wavelength cutoff of the bandpass. The image scale is $0.405''$ pixel$^{-1}$ with a field-of-view of $1.73' \times 1.73'$. The galaxies have angular sizes comparable to the field-of-view and the effect of this on the sky determination is discussed in §3.4.1. The atmospheric transparency was good but not photometric and no flux calibration was attempted. During the four nights, the seeing at K$'$ was $\lesssim 0.8''$. After multiple images were combined, the effective seeing was typically between $1''$ and $1.2''$. The total observing time for any given galaxy was split into about 10 "on" and 10 "off" exposures, alternating every few minutes between the object and a blank patch of sky $\gtrsim 5'$ away. Small random components were added to these offsets to improve flat fielding and to avoid the accumulation of signal in images of faint objects in the "off" images. The total exposure time on each object is listed in Table 1. Additional 5 minute I-band exposures were obtained for six objects using the Las Campanas 1m telescope and a Tektronix 2048 × 2048 CCD.

The reduction of IR array data differs from the reduction of CCD data principally because the background signal at K$'$ is much larger than at optical wavelengths. The sky flux is so overwhelming that the detector must be read out at least every 30 seconds even in the K$'$ band. At a radius of three disk scale lengths a galaxy disk is already 8 magnitudes below the sky level! Therefore, small errors in the sky subtraction can introduce systematic errors and increase the photon noise. To achieve the desired level of precision, images of "blank" sky are used for both flat-fielding and background subtraction. If, as we did during our observations, equal amounts of observing time are spent on object and on sky, then straightforward sky-subtraction will increase the noise by $\sqrt{2}$. This degradation can be mitigated by using a sky image that is the combination of the "off" frames of several objects. Unfortunately, the sky-subtraction noise cannot be reduced arbitrarily in this way, because the sky flux at K$'$ is due both to OH emission in the upper atmosphere and thermal emission in and around the telescope. These two sources illuminate the detector somewhat differently and are time variable. Consequently the total illumination pattern changes shape and scale as the temperature and the atmospheric conditions change. This variability limits the total time over which a sky signal can be combined usefully to less than two hours. We typically achieve a sky flatness good to 0.03% across the frames. Standard IRAF[2] tasks were used to flatten, align, combine, and sky subtract the images.

---

[2]IRAF is distributed by the National Optical Astronomical Observatories, which are operated by AURA, Inc., under contract to the NSF.

# 3 Data Analysis

## 3.1 Fourier Decompositions

Fourier analysis is a natural way to analyze the non-axisymmetric component of the light in a face-on galaxy. It has the advantage over model fitting (see §3.2) that no prior assumption is made about the intrinsic light distribution. Fourier analysis techniques have been used frequently in the past (cf. Grosbøl 1987, Elmegreen et al.1989, Elmegreen et al.1992, RR93, Elmegreen et al.1993, and references therein) and we summarize only briefly the particulars of our implementation. We fix the center of a polar coordinate system at the brightest point of the galaxy in K′, presuming that it accurately represents the bottom of the overall potential well. This choice is justified *a posteriori*, because in their inner portions the sample galaxies only show small asymmetries about the chosen center. Once the center is set, the surface brightness distribution distribution $\mu(R, \varphi)$ is expressed as a Fourier series:

$$\mu(R,\varphi)/\langle\mu(R)\rangle = \sum_{m=1}^{\infty} A_m(R) e^{im[\varphi - \varphi_m(R)]}, \tag{3}$$

where $\varphi$ denotes the azimuthal angle, $m$ the azimuthal wavenumber, and $A_m$ and $\varphi_m$ are the associated Fourier amplitude and phase, respectively. The average surface brightness at radius $R$ is given by $\langle\mu(R)\rangle$.

The analysis sequence consists of several steps. First, we mask all detectable point sources. Second, we re-bin the image onto a $(R, \varphi)$ grid, using 30 bins in radius and 24 bins in azimuth. The rebinning leads to a large speed-up in all subsequent calculations. The minimum and maximum radii for the analysis were set to $\sim$ 6 pixels and 170 pixels, respectively. At small radii ($\lesssim$ 10pix) some interpolation is unavoidable and the data on the polar grid are not fully independent. At larger radii where the number of original pixels per new bin is large, we use the median of the pixels to obtain $\mu(R, \varphi)$. The average and median produce indistinguishable results. The error at each grid point, $\Delta\mu(R, \varphi)$, was taken to be the quadrature sum of the Poisson noise, $\sqrt{(\langle\text{source}\rangle + \langle\text{sky}\rangle)/(\text{number of pixels})}$, and the "flat-fielding error", $\epsilon_{flat} \cdot \langle\text{sky}\rangle$. The flat-fielding error also incorporates sky subtraction errors, which chiefly arise from imperfect flat-fielding. Even though the NICMOS data can be flatfielded extremely well ($\epsilon_{flat} \sim 3 \times 10^{-4}$), flat-fielding errors become dominant at large distances from the galaxy center. In this regime, the errors in Figures 2 and 4 become spatially correlated because the flat-fielding error is assumed to be the same across the whole frame. Lastly, after a $(R, \varphi)$ grid has been established for $\mu$ and $\Delta\mu$, the Fourier amplitudes and phases are determined by a least squares fit to the one-dimensional array of data at each $R$. Solving the least squares problem by means of Normal Equations (cf. Press et al.1986) also yields the variances for each parameter. Systematic uncertainties in this process are described below where we compare the results for an identical analysis of the independent I and K′ images (§3.1.4).

In Figure 2 we plot selected Fourier components for the 18 galaxies. These components were chosen (from the complete Fourier expansion) because they are the only ones that differ significantly from zero in most sample galaxies. The 0th order Fourier amplitude, $A_0$, has no phase associated with it and merely reflects the mean flux in the annulus, $\langle\mu(R)\rangle$. It is plotted as a function of radius in the top panel. For most galaxies the surface brightness profile of the disks is well approximated by an exponential (a straight line in these plots) and we have chosen the exponential scale length as the radial unit of the plots. Our surface photometry typically extends to about three scale-lengths (as determined in Section 3.2 and listed in Table 2). We present an overview of the results for the various $m \geq 1$ Fourier components in the subsequent sections.

### 3.1.1 $m = 1$

Panels two and three in each column of Figure 2 show the $m = 1$ amplitude and phase. The amplitude $A_1(R)$, characterizing the fractional variations in the surface brightness at radius $R$, is related to the fractional offset, $\Delta R/R$, of the isophotes from the center by

$$\frac{\Delta \mu}{\mu} = -\frac{\partial \log \mu}{\partial \log R} \cdot \frac{\Delta R}{R}. \tag{4}$$

For an exponential disk with scale length $R_{exp}$ this implies $\Delta\mu/\mu = R/R_{exp} \cdot \Delta R/R$.

In the inner disk (1 to 2 scale lengths) $A_1$ distortions are generally weak. However, beyond these radii a sizable fraction of galaxies exhibit significant lopsidedness, $A_1 \gtrsim 0.2$. The most prominent examples are IC 2627, NGC 1309, NGC 1325, NGC 1642, NGC 2485, and NGC 6814. An inspection of sky survey plates did not reveal any obvious environmental differences between these lopsided galaxies and the others. One notable exception is NGC 1309 which has an apparent companion visible in our I-band CCD image. A histogram of the lopsidedness in the sample, measured at 2.5 disk scale lengths, is shown in Figure 3. Because the amplitude of the "true" lopsidedness, $\tilde{A}_1(2.5R_{exp})$, is a positive definite quantity, the expectation value of its measurement, $A_1(2.5R_{exp})$, in the presence of an error $\Delta A_1$ is $\sqrt{\tilde{A}_1^2 + \Delta A_1^2}$ (see Section 4.2.2 for a more extensive discussion). We have accounted for this when plotting the histogram. The sample mean of $\tilde{A}_1(2.5R_{exp})$ is 0.14; this value drops to 0.11 if the largest value is excluded. We chose $R = 2.5R_{exp}$ as a radius of comparison because this is the largest radius for which well determined estimates of $\tilde{A}_1$ exist for most the galaxies in the sample.

It is important to note that these results are not affected by an improper choice for the center of our polar coordinate system. If the galaxy isophotes were concentric, but we had chosen the wrong center, then $A_1(R)$ would diverge as $1/R$ for $R \longrightarrow 0$, as long as there is a finite intensity gradient. Furthermore, the asymmetries cannot be a consequence of projection, because that operation is bi-symmetric. Finally, for nearly face-on galaxies the dust extinction on the "near" and "far side" is indistinguishable and will not induce an asymmetry.

### 3.1.2 $m = 2$

The fourth and fifth panels from the top in Figure 2 show the $m = 2$ amplitudes and phases for all sample members. The inclination selection of our sample ($i \lesssim 20°$) limits the apparent ellipticity due to projection effects for circular disks to $A_2(R \sim 2R_{exp}) \lesssim 0.06$. However, the majority of objects exhibit $A_2$'s much larger larger than 0.06 at some radii. Such large amplitudes can arise from either disk intrinsic ellipticity or two-armed spirals. Elliptical disks (or bars) and spirals arms differ in their pitch angle, $\theta \equiv \mathrm{atan}(\partial \log R/\partial \varphi)$, which is expected to be $\sim 90°$ for bar-like distortions, and $0° < \theta < 30°$ for spiral arms. Therefore, the relationship between $\varphi$ and $R$ provides information on the nature of the structure. Not surprisingly, the strongest $m = 2$ elliptical distortions are found in the central bars of the galaxies. The most notable examples in the sample are NGC 600, NGC 1015 and NGC 6814, which have large $A_2$'s for $R < 50''$ at constant $\varphi$ (or $\partial \log R/\partial \varphi = \infty$). At radii beyond this, the strongest features are spiral arms, which have amplitudes of $0.15 \lesssim A_2 \lesssim 0.6$. In many of the galaxies in the sample, the arms wrap continuously through at least $180°$; in IC2617, NGC 1642, NGC 1703, and NGC 6814 they wrap at least $360°$.

### 3.1.3 $m = 3, 4,$ and 6

The bottom three panels in each column of Figure 1 contain the amplitudes of the third, fourth, and sixth Fourier components. Note that the vertical scales in these panels have been stretched by 5/3 relative to the $m = 1$ and $m = 2$ panels because the higher frequency amplitudes are significantly smaller than $A_1$ and $A_2$. Notable exceptions to this are found if strong and thin bars are present (NGC 600 and NGC 1015), then $A_4$ and $A_6$ are comparable to $A_2$. If a bar were infinitely narrow, such that $I(\varphi) = \delta(\varphi) + \delta(\varphi + 180°)$, then all even numbered coefficients, $A_{2n}$ ($n = 1, 2, ...$), would be equal. In practice, bars have finite thickness and the higher order Fourier amplitudes fall off. In some two-armed spirals the non-axisymmetric power is concentrated predominantly in $m = 2$, $\langle A_2 \rangle \gtrsim 2 \langle A_4 \rangle$, which implies that the spiral arms are quite broad (e.g. ESO-436, NGC 1309, NGC 1703, NGC 2466, NGC 2485). However, the two galaxies with the most prominent spiral patterns in the sample, IC 2627 and NGC 6814, also show large $m = 4$ amplitudes, because the strong two arm spirals are narrower than a simple $\cos 2\varphi$ variation. In most cases, terms of order $m > 4$ are small, e.g. $A_6 \ll 0.1$, for nearly all spirals. Three galaxies in the sample, ESO-436, NGC 1376 and NGC 7309 have three-arm spiral patterns that can be easily discerned in the images. These patterns are clearly reflected in the significant $m = 3$ Fourier amplitudes.

### 3.1.4 Comparison of I and K′ Band

Because other K′ data of this type are unavailable, the most direct external error check we can perform is a comparison to our I-band CCD images. However, images in the two colors are not expected to be identical. A multi-color study of M 51 (RR93) demonstrated that the K′ and I band Fourier decompositions can differ, mainly due to obscuration by dust. Nonetheless, many properties such as the overall variations of the $m = 2$ amplitudes and their phases provide a benchmark on the reliability of the K′($2.2\mu$m) data.

We obtained large field, $20' \times 20'$, CCD images in I for six sample members: NGC 1302, NGC 1309, NGC 1376, NGC 1642, NGC 1703 and NGC 2466. We transformed these CCD images to the format of the K′ images and estimated their sky level with the same procedure used for the NICMOS data. We then performed a Fourier decomposition using the same grid as that used for K′ data. The resulting comparison for the six objects is shown in Figure 4. The top panel contains the I-K′ color, subject to an arbitrary offset. All galaxies become slightly bluer with increasing radius for radii $\lesssim 40''$. The interpretation of this color gradient in terms of population and dust extinction gradients is beyond the scope of the present paper. At radii $\gtrsim 40''$, the uncertainties in the sky level increase beyond $\sim 0.1 - 0.2$ mag. However, for all galaxies the colors at large radii are consistent with an extrapolation of the color gradient at smaller radii. We take this as an indication that we have conservatively estimated our sky errors. For most of the galaxies, the $m = 1$ and 2 amplitudes and phases are consistent over most of the radial range. Wherever the Fourier phases differ by large amounts, differ either by $360°/m$ or in regions where the associated Fourier amplitude is very small. Note that discrepant amplitudes do not primarily occur in the outer low signal-to-noise portion of the profile, but rather in the inner, well measured portion of the profile. These differences are presumably caused by dust. Note in particular that the agreement of the $m = 1$ distortion between I and K′ shows that these asymmetries are neither due to gradients in the sky background nor due to asymmetric dust extinction. For $20'' < R < 60''$, $m = 2$ is almost always larger in the K′ data than in the I data. This is consistent with the results for M51 (RR93), where a significant fraction of the I-band $m = 2$ amplitude is suppressed by dust along the ridges of the arms.

The principal function of the I-band images is to test for distortions in the K′ images caused by a poor sky level determination whenever a galaxy fills the field-of-view. For this test we compared the luminosity profiles derived from the trimmed I images with the results from the untrimmed images, for which the sky level was determined from the much larger field-of-view. In all cases the two sets of profiles agree within the error bars. Note that this test does not depend on the absolute sky level (which is much higher in K′ ), but only on the degree to which the galaxies fill the field-of-view.

The general agreement between the K′ and I results, and between the results from large and small field-of-view I images, substantiates the assigned K′ error bars.

## 3.2 Model Fitting

Because the $m = 2$ amplitude from the Fourier analysis reflects a mixture of inclination effects, bulge and disk ellipticity, and spiral arms, it cannot be used to place unambiguous constraints on the shapes of disks. Without further information these various effects cannot be untangled rigorously. We make progress by concentrating on the ellipticity of the disk at a radius-independent, but arbitrary, position angle. By fitting a specific model to the images averages radially over the distortions created by the spiral arms. As we derive in Section 4.2.1, for radii greater than $\sim R_{exp}$ the isophotes of a disk residing in a non-rotating logarithmic potential of ellipticity $\epsilon_\Phi$ (in the disk plane) have an ellipticity, $\epsilon_{iso}(R) = \epsilon_\Phi(1 + R_{exp}/R)$ (Franx and de Zeeuw, 1992). The first step in this analysis is to derive the radial range over which the disk dominates the total light ($\mu_{disk} > 2\mu_{bulge}$) by fitting an axisymmetric bulge-disk model to the whole image. Then we estimate $\epsilon_\Phi$ by fitting a disk model with an ellipticity profile given by $\epsilon_{iso}(R)$ over a restricted radial range of the image (see Appendix A for details). We adopted this procedure, rather that fitting simultaneously non-axisymmetric bulges and disks of independent ellipticities and position angles. This simultaneous fit often yielded strongly distorted bulges and disks oriented at right angles. The superposition of a perpendicularly oriented bulge and disk produces a much less flattened overall image than each component separately: a presumably unphysical result. For each galaxy our fit provides an estimate of the scale length, $R_{exp}$, the ellipticity of the potential, $\epsilon_\Phi$, and position angle, $PA$, and their associated uncertainties. These results are presented in Table 2. With a few exceptions (e.g. NGC 1302 and NGC 2718), the values for $\epsilon_\Phi$ are found to be $\lesssim 0.1$.

Because a perfectly open (a pitch angle of 90°) spiral arm is indistinguishable from disk ellipticity, we use simulated data, constructed from a perfectly axisymmetric and exponential disk with a superimposed spiral, to study the coupling between the arms and the disk ellipticity. The simulated spiral winds logarithmically with a pitch angle of 20°, and extends from one to three disk scale lengths. These parameters are characteristic of the spiral pattern found in our sample. We set $A_2$ for the simulated galaxy to match the one observed in NGC 1703 and assumed $A_4(R) = A_2(R)/2$. The simulated galaxy image is shown in Figure 5, and seems to mimic a typical grand design spiral. We then add noise and sky errors to this image that are matched to the noise properties of our data. Finally, we determine the disk's ellipticity using the same fitting procedure as before and find $\epsilon = 0.03$ for this simulated spiral. This result suggests that although spiral arms do contribute significantly to the measured disk ellipticity, they do not dominate the signal in the $\epsilon_\Phi$ estimates for most sample members. In our final analysis we treat this interplay between disk ellipticity and spiral arms by quadratically adding an extra error term of $\Delta \epsilon_{Sp} = 0.03$.

## 4 Discussion

### 4.1 Surface Brightness vs. Surface Mass Density

To assess the implications of the K′(2.2$\mu$m) Fourier decomposition and of the model fitting, it is important to understand the degree to which surface brightness traces the surface mass density. This mapping is complicated by dust extinction and by spatial variations in the stellar mass to light ratio. The first effect is largely eliminated by observing in the K′ band, where the dust optical depth is reduced by a factor 6 compared to $I(0.8\mu)$ (e.g. RR93). But because a population of young stars is brighter at all wavelengths than corresponding old stars, the common lore that "red light arises from old stars" may be dangerously misleading.

There are two ways of estimating the contribution of young stars to the surface brightness enhancement in the spiral arms. First, because young stellar populations are both brighter and bluer than older ones, one can compare the arm/inter-arm *color change* to the arm/inter-arm *brightness contrast* (Schweizer, 1976; Jensen, Talbot and Dufour, 1981; Elmegreen and Elmegreen, 1984; Kennicutt and Edgar, 1986). These authors demonstrate that the color change across the spiral arm is too small to explain the brightness contrast merely through azimuthal mass-to-light ratio variations; consequently they argue for a substantial enhancement of the stellar surface mass density in the spiral arms of grand design spirals. Alternatively, one can estimate the contribution of luminous young stars in the IR directly. RR93 attempted to do so by measuring the CO band-head strength in the disk of M51 and concluded that although young stars may contribute significantly ($\sim 20\%$) to the K-band light in the arm crests, they cannot be responsible for the bulk of the light from the spirals arms. The situation for M 33 appears similar. Regan and Wilson (1993) obtained B and V band photometry for part of the southern spiral arm of M33 and detected red supergiants in some of the stellar associations. However, even in those regions, the RSGs comprise only 10% of the K band flux (assuming that they are 2 magnitudes redder in V-K than the underlying disk). Therefore, at least where one can study the spiral arm population in detail, the light from young stars contributes only between 10 and 20% of the total K-band light.

Because the galaxies in our sample are too distant for a detailed study of their stellar content, we supplement the previous discussion by a third, independent argument: a brief discussion of the theoretically expected "age distribution" of the K-band light. As detailed in Appendix C, we use the stellar population models of Charlot and Bruzual (1991) and simple assumptions about the star formation histories of galaxies (e.g. Kennicutt 1983; Kennicutt, Tamblyn, and Congdon 1994) to estimate the fraction of the total K light contributed by stars younger than age $t_y$, $L_K(<t_y)/L_K(<t_0)$, where $t_0$ is the current age of the Universe. This age distribution is shown in Figure 6 and it implies that 8% of the IR light arises from stars with $t \lesssim t_{orbit} \sim 10^8$ yrs and 18% from stars with $t \lesssim 10 t_{orbit}$. Because stars phase mix or drift out of the spiral pattern in $\lesssim 10 t_{orbit}$, these models imply that the upper limit on the total nonaxisymmetric power (all Fourier components except $A_0$ summed) that could arise from azimuthal population variations is 20%. However, this line of reasoning also suggests that about half of the light from stars with $t < 10 t_{orbit}$ should come from stars with $t < t_{orbit}$. Because stars with ages substantially less than $t_{orbit}$ have not drifted far from their birth place, the azimuthal population variations arising from these stars should be sharply defined. When the distribution is sharply defined in azimuth (as in the bar of NGC 1015) then $A_2 = A_4 = A_6$. If the spread in azimuth is less than $2\pi/m_n$, we expect this equality to hold for all even $A_m$, $m \ll m_n$. However, $A_4$ is observed to be generally considerably smaller than $A_2$ (see §3.1.3), in conflict with these model predictions. The discrepancy may be removed if the models are overestimating the K' luminosity of very young stars. Nevertheless, the results from these calculations suggest an upper limit to the contamination of the total IR luminosity from young stars of 20%.

The empirical evidence and the model predictions suggest that the assumption of mass tracing light at K' is valid within $\sim 20\%$. The regime in which it is most likely to fail is at large azimuthal wave numbers ($m > 2$) and small amplitudes ($A_m \lesssim 0.1$).

## 4.2 How Elliptical Are Galaxy Disks?

Are galaxy disks triaxial? Or equivalently, are they elliptical when viewed perpendicular to the disk plane? We cannot answer these questions as straightforwardly as we can measure lopsidedness, because two-armed spirals and projection effects affect measurements of triaxiality. Therefore, we focus on a specific form of disk ellipticity: a disk residing in a logarithmic (halo) potential of constant fractional distortion and without figure rotation. The choice of a nonaxisymmetric and logarithmic potential is motivated by the approximately flat observed rotation curves of spiral galaxies out to several optical radii (e.g. Begeman 1987 and references therein). Setting the pattern speed to zero is motivated by Occam's razor and by the fact that it avoids dealing with resonances. These assumptions do not necessarily conflict with the existence of wound spirals arms of finite pattern speed, which may occur in a triaxial potential.

### 4.2.1 Isophote Shapes and the Gravitational Potential

With the assumption that most stars at $R \gtrsim R_{exp}$ are on loop orbits one can derive a unique relation between the shape of the potential and the shape of face-on isophotes (see Franx and de Zeeuw 1992). Starting with a flattened potential with (small) ellipticity $\epsilon_\Phi$ in the equatorial plane and a characteristic circular velocity $v_c$,

$$\Phi(R, \varphi) \equiv \frac{v_c^2}{2} \ln R^2 - \frac{\epsilon_\Phi}{2} \frac{v_c^2}{2} \cos 2\varphi + \text{const.}, \tag{5}$$

and applying first order perturbation theory (as outlined in Binney and Tremaine 1987, pg. 146) one can obtain the set of closed loop orbits at the mean radius $R_0$

$$R(\varphi) = R_0 \left(1 - \frac{\epsilon_\Phi}{2} \cos 2\varphi\right), \tag{6}$$

$$v_\varphi(\varphi) = v_c \left(1 + \epsilon_\Phi \cos 2\varphi\right), \tag{7}$$

and

$$v_R(\varphi) = -v_c \epsilon_\Phi \sin 2\varphi. \tag{8}$$

These orbits have the same ellipticity as the potential, but are elongated ($\varphi_{maj}^{orb} = 90°$) perpendicularly to the major axis of the potential ($\varphi_{maj}^{pot} = 0°$). The velocity along the orbit varies by twice $\epsilon_\Phi$, with $v_\varphi(max)$ occurring at $R(min)$, while the extrema in $v_R$ are off-set by 45°. An ensemble of stars on such orbits must satisfy the continuity equation

$$\frac{\partial}{\partial R}\left[R \cdot \mu(R, \varphi) \cdot v_R(\varphi)\right] + \frac{\partial}{\partial \varphi}\left[\mu(R, \varphi) \cdot v_\varphi(\varphi)\right] = 0, \tag{9}$$

where $\mu$ denotes the surface mass density. If the radial surface brightness profile is an exponential, we can write $\mu(R, \varphi)$ as

$$\mu(R, \varphi) = \mu_0 \exp\left[-\frac{R}{R_{exp}}\left(1 - \frac{\epsilon_{iso}(R)}{2} \cos 2\varphi\right)\right], \tag{10}$$

where $\mu_0$ is the central surface brightness and $\epsilon_{iso}$ is the ellipticity of the isophote. We obtain from Eqs. (6) – (10)

$$\epsilon_{iso}(R) = \epsilon_\Phi \left(1 + \frac{R_{exp}}{R}\right). \tag{11}$$

Eq. (11) demonstrates that the disk isophotes are always more distorted than the potential, *e.g.* by a factor of 1.5 at two scale lengths. The divergence of $\epsilon_{iso}(R)$ as $R$ approaches zero is of no practical consequence because the assumption that stars are on loop orbits fails for $R \ll R_{exp}$ (e.g. Kuijken 1993); we only consider the regime for which $R \gtrsim R_{exp}$.

So far we have only considered streamlines of closed orbits, such as those of a kinematic tracer with negligible radial velocity dispersion (e.g. H I). Stars, however, have a non-negligible dispersion, and move on orbits that are "sired" by the closed parent orbit. The general form of a non-closed orbit in the above potential is

$$R(t) = R_0 \left(1 - \frac{\epsilon_\Phi}{2} \cos 2\Omega t + C \cos \sqrt{2}\Omega t\right) \qquad (12)$$

(Binney and Tremaine 1987), where $\Omega$ is the angular frequency of the unperturbed orbit. As evident from Equation 12, the stars oscillate with an irrational frequency about the location of the parent orbit. To first order, the amplitude $C$ of this oscillation, which can be derived by differentiating Eq. 12, is given by $C^2 = \langle v_R^2 \rangle / v_c^2$. In the solar neighborhood $C \sim 0.2$ and hence is much larger than the expected contribution from $\epsilon_\Phi/2$. An ensemble of such orbits forms an outer envelope of ellipticity $\epsilon_{outer} = 1 - (1 - \frac{\epsilon_\Phi}{2} + C)/(1 + \frac{\epsilon_\Phi}{2} + C)$ and an inner envelope of $\epsilon_{inner} = 1 - (1 - \frac{\epsilon_\Phi}{2} - C)/(1 + \frac{\epsilon_\Phi}{2} - C)$. Accounting for the greater weight of the outer parts to the observed isophote (due to the slower azimuthal velocities at apocenter) and averaging the inner and outer edge of such an annulus, one finds that the ellipticity of the isophotes only changes by $-\epsilon_\Phi C^2$ compared to the closed orbits, with all lower order terms vanishing. Therefore, the isophote shapes are independent of the stellar velocity dispersion in the limit $v_R \ll v_c$. We will use this approximation throughout the paper and it enables us to constrain the potential shape simply from the isophote shapes.

### 4.2.2 Estimating the Distribution of $\epsilon_\Phi^I$

In Section 3.2 we described how we estimate the *observed* disk ellipticity $\epsilon_\Phi$ and its uncertainties for each of the sample members. If our galaxies were perfectly face-on and the ellipticity error were negligible, the results from Section 3.2 would give an immediate observational estimate of the *intrinsic* ellipticity distribution, $P(\epsilon_\Phi^I)$, for the sample.

In practice our estimate of $P(\epsilon_\Phi^I)$ must account for two complicating effects: ambiguities in the deprojection and errors in ellipticity measurement. When the disk has a small, but finite, inclination, there is a range of intrinsic ellipticities that can result in the observed ellipticity. If we denote the projected position angle of the intrinsic disk major axis as $PA_{maj}$ and the projected position angle of the disk normal direction as $PA_{\hat{n}}$, then this projected ellipticity depends on $PA_{maj-\hat{n}} \equiv PA_{maj} - PA_{\hat{n}}$. Because our inclination estimates come from single beam H I measurements, we have no prior knowledge of $PA_i$; we therefore must assume that $PA_{maj-\hat{n}}$ is uniformly distributed. The ellipticity errors, $\Delta\epsilon$ enter the estimate of $P(\epsilon_\Phi)$ in two ways: (1) $\epsilon_\Phi^I$ can obviously not be measured perfectly, and (2) measurement errors lead to a biased estimate of $\epsilon_\Phi^I$ because ellipticities are positive definite quantities. Fortunately, analogous probability distributions for positive definite quantities have been derived in other astrophysical contexts (e.g. Wardle and Kronenberg 1977). We describe the application of those distributions in detail in Appendix B.

In the subsequent sections we describe how to estimate $P(\epsilon_\Phi^I)$ from our sample in both a parametric and a non-parametric fashion. In the first approach we do not need to adopt a functional form for $P(\epsilon_\Phi^I)$. This is an important asset because we have no physical arguments for any particular form. However, such a technique will not recover the "true" distribution even for an infinite number of measurements. In contrast, we *can* obtain an unbiased estimate if we adopt a parametric functional form for the $P(\epsilon_\Phi^I)$. Our philosophy here is to first estimate $P(\epsilon_\Phi^I)$ non-parametrically and then to use this result to choose a functional form for the parametric estimate.

*a) Non-Parametric Estimates*

As shown in Appendix B, one can determine from the observed parameters for each galaxy $j$, the probability distribution of intrinsic ellipticities, $P_j(\epsilon_\Phi^I)$, $\cos i$, $\epsilon_\Phi$, $\Delta \epsilon_\Phi^I$. This estimate is based on the assumption that the angle specifying the disk orientation at given inclination is distributed uniformly. The probability distribution for the entire sample is then

$$P(\epsilon_\Phi^I) \equiv \frac{1}{N} \sum_{j=1,N} P_j(\epsilon_\Phi^I) \tag{13}$$

This distribution is shown as the thick solid line in Figure 7. We estimate the 95% confidence region for this distribution *due to our finite sample size* by means of "bootstrapping" (e.g. Press *et al.*, see Appendix B). These estimates are shown as the two dashed lines enveloping our estimate of $P(\epsilon_\Phi^I)$.

This estimate will never converge to the true answer. Consider for example an ensemble of axisymmetric disks, $P_{true}(\epsilon_\Phi^I) = \delta(\epsilon_\Phi^I)$, seen at small but finite inclinations. Then all disks will project into slightly elliptical shapes. If we have no prior knowledge about $P_{true}(\epsilon_\Phi^I)$, we can never exclude the possibility that some disks are truly elliptical. Finite measurement errors lead to a similar effect. Therefore, our derived distribution $P(\epsilon_\Phi^I)$ should be compared to the distribution, $P_0(\epsilon_\Phi^I)$, that would result if we observed a set of perfectly round disks with inclinations and uncertainties corresponding to the disks in the sample. The distribution $P_0$ and its 95% confidence regions are shown as the dotted lines in Figure 7. $P_0$ can be thought of as a "point-spread function" because it reflects the estimate of $P(\epsilon_\Phi^I)$ for an input $\delta$-function.

The observed $P(\epsilon_\Phi^I)$ appears broader than $P_0(\epsilon_\Phi^I)$, even when considering the range of the confidence limits. The median $\epsilon_\Phi^I$ for axisymmetric disks is 0.025 but is 0.07 for our sample galaxies, showing that *the underlying stellar disks in these galaxies are slightly elliptical*. However, these non-parametric estimates do not provide a good measure for a "characteristic" ellipticity, $\epsilon_0$, and complicate the presentation of confidence limits.

*b) Parametric Likelihood Estimates*

The observed $P(\epsilon_\Phi^I)$, displayed in Figure 7, prompted us to chose an exponential as the parametric trial function for $P(\epsilon_\Phi^I)$:

$$P(\epsilon_\Phi^I, \epsilon_0) = \frac{1}{\epsilon_0} \exp\left[-\frac{\epsilon_\Phi^I}{\epsilon_0}\right]. \tag{14}$$

Our task is now to determine the most likely value of $\epsilon_0$. Note that the mean and the variance of this distribution are both equal to $\epsilon_0$. For each assumed $\epsilon_\Phi^I$, inclination, and $PA_{maj-\hat{n}}$, we can determine the expected ellipticity for galaxy $j$ in the noiseless case, $\epsilon_{no-noise}$. In the presence of measurement error, $\Delta \epsilon$, the distribution of measured $\epsilon_\Phi$ for galaxy $j$ is given by (Wardle and Kronenberg 1977)

$$\tilde{P}_j(\epsilon_\Phi)|_{\epsilon_\Phi^I, \cos_j(i), PA_{maj-\hat{n}}} = \frac{\epsilon_\Phi}{\Delta \epsilon^2} I_0\left(\frac{\epsilon_\Phi \epsilon_{no-noise}}{\Delta \epsilon^2}\right) \exp\left(-\frac{\epsilon_\Phi^2 + \epsilon_{no-noise}^2}{2\Delta \epsilon^2}\right), \tag{15}$$

where $I_0$ is the modified Bessel function. Integrating $\tilde{P}_j$ over all (unknown) angles $PA_{maj-\hat{n}}$ and all $\epsilon_\Phi^I$ yields the probability distribution $P_j(\epsilon_{\Phi,j}, \epsilon_0)$ of measuring $\epsilon_\Phi$ in galaxy $j$, which is now only a function of $\epsilon_0$ and the data. We proceed by defining the likelihood, $\mathcal{L}$, for the parameter $\epsilon_0$ to be

$$\mathcal{L}(\epsilon_0) \equiv \sum_{j=1..N} \ln\left(P_j(\epsilon_{\Phi,j}, \epsilon_0)\right) - N, \tag{16}$$

where the sum is taken over all $N$ sample galaxies and $\epsilon_{\Phi,j}$ is the measured value for the $j^{th}$ galaxy. Once the best fit $\epsilon_0$ is found, we estimate its variance by finding the parameter region whose likelihood is within unity of the best fit value.

This analysis is illustrated in Figure 8, where we show a histogram of the measured $\epsilon_\Phi$ along with the expected distributions of measurements for $\epsilon_0 = 0.025, 0.045,$ and $0.075$. Maximizing $\mathcal{L}(\epsilon_0)$ yields the following results: (1) if we assume a spurious ellipticity due to spiral arms of 0.03 (see §3.2), which is added in quadrature to the measurement error, we find $\epsilon_0 = 0.045^{+0.03}_{-0.02}$, (2) if we do not include this source of error, then more of the observed ellipticity becomes attributable to the underlying disk ellipticity and $\epsilon_0 = 0.055^{+0.03}_{-0.02}$, (3) if we assume that our estimate of $\cos(i)_{HI}$ from the linewidths is only an upper limit, and that in reality the galaxies are distributed uniformly between $\cos(i)_{HI}$ and $\cos(i) = 1$, then less of the observed distortions is due to inclination effects and $\epsilon_0$ increases to $0.055^{+0.04}_{-0.02}$, and (4) if we assume the same conditions as in (1), but eliminate the two most strongly distorted galaxies (NGC 1302 and NGC 2718), we find $\epsilon_0 = 0.030^{+0.03}_{-0.015}$. *Therefore, we conclude that the galaxy disks in our sample have small but finite ellipticity.* The characteristic ellipticity of the potential, within the disk plane, is $\epsilon_0 = 0.045^{+0.03}_{-0.02}$, and this result is quite insensitive to the details of the statistical treatment. The potential ellipticity, $\epsilon_\Phi = 0.1$, inferred for the Milky Way by Kuijken and Tremaine (1994) is somewhat higher than the mean found here, but still consistent with our distribution for $\epsilon_\Phi$ determined from external galaxies.

It is worth stressing again that the biggest source of uncertainty in this analysis is the cross-talk between two-arm spirals and disk ellipticity. Even though we made an attempt to account for this through simulations, it is conceivable that spiral arm effects are bigger than we estimated. In this case, our $\epsilon_0$ estimates still can be considered as a conservative upper limit on the triaxiality of the potential.

### 4.2.3 Kinematic Implications

These ellipticity estimates have kinematic consequences from which we can make predictions about the radial and azimuthal motions in disk galaxies. Elliptical streamlines arising in triaxial disks will result in nonzero radial velocity components for the local standard of rest (LSR) at all points along the orbit except at apocenter and pericenter. The LSR is defined to be co-moving with the net streaming of an ensemble of stars. Using Equation 8, the *rms* (denoted by $\langle ... \rangle$) radial velocity of the LSR, averaged over all azimuthal angles is

$$\langle \frac{v_R}{v_c} \rangle = \frac{\epsilon_\Phi}{\sqrt{2}}. \tag{17}$$

Using characteristic values for our sample, $v_c = 200$ km s$^{-1}$ and $\epsilon_0 = 0.045$, we find that along a streamline the radial velocity with respect to the galaxy center varies from $\sim -9$km s$^{-1}$ to $\sim 9$km s$^{-1}$ with $\langle v_R \rangle = 6.4$ km s$^{-1}$ .

This implies that the value inferred for the radial motion of the LSR in the solar neighborhood (14km s$^{-1}$ , Blitz and Spergel 1990) is consistent with expectations based on the observed elliptical disks for a "typical" spiral galaxy. However, our data are equally consistent with the non-detection ($v_R = -1 \pm 9$km s$^{-1}$ ) found by Kuijken and Tremaine (1994). The elliptical streamlines also lead to variations of the azimuthal velocity along the orbit (see Eq. 7). The fractional variance of this velocity is given by

$$\langle \frac{v_\varphi}{v_c} - 1 \rangle = \frac{\epsilon_\Phi}{\sqrt{2}}. \tag{18}$$

As pointed out by Franx and de Zeeuw (1993), this variation contributes to the scatter in the TF relation — even if the true inclinations can be determined perfectly. Because the measured

H I linewidth is presumed to be proportional to $2v_{tangential}(\varphi)/\sin i$, it will vary among identical galaxies seen at the same inclination but at different angles $\varphi$. This effect leads to a scatter (in magnitudes) of at least

$$\Delta M = \frac{7.8}{\ln(10)} \langle \frac{v_\varphi}{v_c} - 1 \rangle, \tag{19}$$

for the TF relationship given by Pierce and Tully (1992). A characteristic ellipticity of $\epsilon_0 = 0.045$ would then produce a scatter of 0.15 magnitudes. However, an immediate comparison to the scatter in the TF relation from observed samples of galaxies is complex. The K$'$(2.2$\mu$m) photometry and H I linewidths presumably probe triaxiality at different radii. While our data probe typically $3-9$ kpc, the H I flux arises typically from radii larger by a factor of two. (S. Rao, *priv. comm.*). In addition, most TF observations are intended for distance estimates, therefore the samples are rarely statistically complete and may be biased against morphological or kinematic peculiarities.

### 4.3 How Lopsided are Galaxy Disks?

It is common lore that some galaxies appear lopsided, e.g. Sandage (1961) commented in the "Hubble Atlas" on the asymmetry of M101, NGC 1637, and others. Nonetheless, the study of Baldwin, Lynden-Bell, and Sancisi (1980) appears to be the only systematic study of such asymmetries in the literature. These authors studied lopsidedness in the H I distributions of galaxies, discussed excitation mechanisms, and estimated the longevity of such modes. The data presented in Figures 1 and 3 show that lopsidedness in stellar disks is also common. Note that $m = 1$ amplitudes of $\gtrsim 0.2$ are found in the outer parts of about one third of our sample. Regardless of the cause or longevity of such $m = 1$ distortions, they, like disk ellipticity, will lead to a radial velocity of the LSR in these galaxies. If we assume, in analogy to the $m = 2$ case, that to first order the closed streamlines at radius $R_0$ have the shape

$$R(\varphi) = R_0 \left( 1 - \frac{\epsilon_{lop}}{2} \cos \varphi \right), \tag{20}$$

then the azimuthal and radial velocities have the form

$$v_\varphi(\varphi) = v_c \left( 1 + \frac{\epsilon_{lop}}{2} \cos \varphi \right). \tag{21}$$

and

$$v_R(\varphi) = -v_c \frac{\epsilon_{lop}}{2} \sin \varphi, \tag{22}$$

respectively. Applying the continuity equation (Eq. 9) to these orbits shows that $m = 1$ streamlines have the same fractional offset, $\epsilon_{lop}$, as the isophotes. For an exponential brightness profile this offset is related to the Fourier amplitude by

$$\epsilon_{lop} = 2\tilde{A}_1 \frac{R}{R_{exp}}. \tag{23}$$

Averaging over Eq. 22 and using Eq. 23, we now express the expected *rms* motion of the LSR at radius $R$ in terms of the observables $\tilde{A}_1$ and $R_{exp}$,

$$\langle \frac{v_R}{v_c} \rangle = \frac{\tilde{A}_1 R_{exp}}{\sqrt{2} R}. \tag{24}$$

We rewrite the equation for $\langle v_R \rangle$ using characteristic values in our sample

$$\langle v_R \rangle = 7.4 \text{ km s}^{-1} \left[ \frac{v_c}{200 \text{ km s}^{-1}} \right] \left[ \frac{\tilde{A}_1}{0.11} \right] \left[ \frac{2.5 R_{exp}}{R} \right]. \tag{25}$$

*Consequently, the expected coherent radial motions from lopsided and elliptical distortions are of the same magnitude* and both are comparable to the value inferred for the solar neighborhood (e.g. Blitz and Spergel 1990).

These dynamical inferences are valid regardless of the origin and the longevity of the $m = 1$ distortions, provided the distortion persists longer than one orbital time. However, the question of the longevity of lopsidedness is both important and puzzling. As first discussed by Baldwin *et al.* (1980), if lopsidedness is due to elliptic orbits in an *axisymmetric* logarithmic potential, any initial azimuthal alignment of the orbit apocenters between the radii of $R_1$ and $R_2$ would wind into a leading spiral at a rate of

$$\Omega_{wind} = \left(\sqrt{2} - 1\right) v_c \left(\frac{1}{R_{inner}} - \frac{1}{R_{outer}}\right). \tag{26}$$

Using values characteristic for our sample, $v_c = 200$ km s$^{-1}$, $R_{outer} = 1.5 R_{inner}$ and $R_{outer} = 8$kpc, the "wind-up time," $2\pi \Omega_{wind}^{-1}$, is only about $10^9$ yrs. This estimate of the lifetime in conjunction with our observation of significant lopsidedness in about 1/3 of our sample suggests that the generation of lopsidedness has a typical time-scale of $3 \times 10^9$ yrs. Alternatively, we could suppose that disk stars merely respond to a potential distortion of the form $\Phi(R, \varphi) = \Phi(R) + (\epsilon_{lop}/2)\cos\varphi$. Then all closed orbits are aligned eccentric ellipses of the form given in Eq. (13) and there is no winding problem. However, as in the $m = 2$ case, the stellar orbits "oppose" the potential they reside in and cannot create such distortions self-consistently. Therefore the winding problem is merely replaced by the question of how to create a lopsided potential.

## 4.4 How Strong are Spiral Arms?

Two empirical statements made in Section 3.2 about the appearance of two-armed spirals at K' bear repeating : (1) they are usually coherent features, extending over a factor of about two in radius, and often winding through at least 180°, and (2) they are azimuthally broad features, in the sense that $A_2$ is the dominant even Fourier amplitude. Both of these properties are consistent with the hypothesis that the arms chiefly reflect variations in the stellar surface mass density (see Section 3.1). These IR observations are in support of optical studies (Schweizer, 1976; Jensen *et al.*1981; Elmegreen and Elmegreen, 1984), which show insufficient color change across the arms to explain the intensity variations merely with stellar population changes. In Figure 9 we give a more quantitative picture of the spiral arm strength. In this Figure we plot the fractional variation $\Delta(R) \equiv I_{max}(R)/I_{min}(R) - 1$, where $I_{max}$ and $I_{min}$ are the maximum and minimum value of the surface brightness in a model image. To focus on features with two-fold symmetry we constructed a model image consisting only of the $m = 0, 2, 4, 6$ Fourier components in the observed images. The uncertainties in Figure 9 were calculated from the uncertainties in the Fourier components. As discussed earlier, these intensity variations in azimuth reflect a mix of projection effects, disk ellipticity and spiral arms. However, projection effects contribute only $(1 - \cos i)\frac{R}{R_{exp}}$ and (disk) ellipticity accounts for only $\epsilon_\Phi(1 + \frac{R_{exp}}{R})\frac{R}{R_{exp}}$; both effects cause contributions of $\Delta(R) \lesssim 0.2$. Since the variations displayed in Figure 1 are very often larger than these effects, they are most likely due to bars (at small radii) and spiral arms (at larger radii). Although we are not able to devise a physically well motivated definition of "spiral arm strength", a comparison of Figure 9 with Figure 2 (where one can check whether the $m = 2$ phase winds continuously) suggests that nearly half of the galaxies exhibit spiral arms with an arm-inter arm contrast, $\Delta(R)$, of about unity.

If these luminosity variations indeed reflect mass variations, as the available evidence suggests, these spiral arms will have observable dynamical implications. We are able to make straightforward predictions about non-circular motions for elliptical or lopsided disks because (as we showed in §4.2.1) the K′ light traces the streamlines in a nonaxisymmetric potential. But even if we assume that the observed K′ surface brightness $\mu(R,\varphi)$ perfectly traces the stellar surface mass density in the disk, $\Sigma(R,\varphi)$, and that the spiral pattern is stationary in some reference frame rotating at $\Omega_p$, the analysis of the spiral arms is more complex than that of either ellipticity or lopsidedness (e.g. Contopoulos and Grosbøl, 1988, Grosbøl, 1993).

Here we restrict ourselves to a brief discussion of the non-axisymmetric forces arising from the spiral arms, assuming that light traces mass. The Fourier decomposition of our images provides a convenient separation of the axisymmetric, $A_0(R)$, and nonaxisymmetric, $A_{m>0}(R)$, components of the surface mass density. Using only $A_0(R)$, we can determine the radial, $f_R^0(R)$, and azimuthal, $f_\varphi^0$, forces (which vanish identically) from the axisymmetric component of an observed galaxy. We calculate these forces by assuming a two-dimensional mass distribution and summing directly over the bins in our polar data grid. Similarly, we calculate $f_R^m(R)$ and $f_\varphi^m(R)$ from $\mu(R,\varphi)$. To focus on the effect of two armed spirals we include only the $m = 2$ and $m = 4$ components in our force calculation. The normalized forces arising from the nonaxisymmetric part, are defined by

$$F_R^m(R,\varphi) \equiv \frac{f_R^m(R,\varphi) - f_R^0(R)}{f_R^0(R)} \tag{27}$$

and

$$F_\varphi^m(R,\varphi) \equiv \frac{f_\varphi^m(R,\varphi)}{f_R^0(R)}. \tag{28}$$

To account for centripetal force from an axisymmetric halo, the denominator should be replaced by $f_R^0(R) + f_R^{halo}(R)$. Here we calculate upper limits on $F_R$ by adopting $f_R^{halo}(R) = 0$. The nonaxisymmetric *rms* normalized forces, for $m = 2$ and 4, for two galaxies with strong spiral patterns, IC 2627 and NGC 6814, are between 5 and 15%. A similar analysis of I-band data for M81 (Visser 1980), which is a weak-armed spiral (Schweizer 1976), produced an estimate of $\sim 5\%$. This demonstrates that even though the mass density contrast across a spiral arm may be near unity, the force perturbations are a much smaller fraction of the total forces. Nevertheless, the forces are not insignificantly small and may pose problems for the standard linearity assumption in applications of the spiral density wave theory.

## 5 Conclusions

The study presented here shows that galaxy disks exhibit a bewildering variety of shapes even if one observes in the near-IR to minimize distortions due to dust and azimuthal mass-to-light ratio variations. We used Fourier techniques and model fitting to study the nonaxisymmetric component of the stellar disks. These techniques enabled us to decompose the nonaxisymmetric part of the K′(2.2$\mu$m) surface brightness into (a) disk ellipticity, (b) lopsidedness and (c) two arm spirals. The basic result from our analysis is that all three types of nonaxisymmetries are common. Our study indicates that distortions in the K′(2.2$\mu$m) light principally reflect distortions in the stellar surface mass density, rather than variations in the mass-to-light ratio. This mapping of light to mass can be tested further in dynamical studies or by comparing the current rate and distribution of star formation to the K′(2.2$\mu$m) light.

Estimates of the dynamical effects arising from the various nonaxisymmetric components showed that all three types of distortions are comparably important, in the sense that they all induce velocities (in stars or gas) which differ typically $\sim 3 - 6\%$ from those found in "equivalent" axisymmetric galaxies. The peak values of the nonaxisymmetric motions are expected to be higher by factors of at least two and should be easily detectable in velocity maps of ionized or neutral gas.

If we assume that the Milky Way is a "typical galaxy", our results can provide a *prior* probability on local distortions. Even though we have not considered distortions with finite pattern speeds, it seems that all models suggested so far (Blitz and Spergel, 1990; Kuijken 1991; Kuijken and Tremaine, 1994) have comparable prior likelihood. The one form of distortion that deserves more attention is massive spirals. Their modeling in the MW, however, is considerably more complex, especially since the spiral arms in other galaxies exhibit such a wide variety of morphologies.

Our results confirm the assertion by Franx and de Zeeuw (1992) that non-axisymmetries may contribute significantly to the observed scatter in linewidth-luminosity relations for disk galaxies. Our results are most immediately applicable to studies where the kinematics are taken from ionized gas (e.g. Matthewson *et al.*1992, Schommer *et al.*1993), because similar radii are probed with these kinematics and our photometry. In these studies, disk ellipticities are expected to contribute $\sim 0.15$mag scatter.

The present results can be tested further in a variety of ways: the most obvious way is to check whether the non-circular motions suggested in §4 are borne out in the velocity fields of external galaxies. The most suitable kinematic data set would be HII velocity fields derived from Fabry-Perot imaging and HI 21-cm maps. For the first type of data it is often impossible to obtain a complete 2-D velocity filed, because the HII emission is very clumpy. For the second type of data it is difficult to obtain sufficiently high spatial resolution. Such a data model comparison also requires a more detailed prediction of the velocity field than provided in §4. For lopsided galaxies the most straightforward dynamical test is a search for a shift of the "systemic" velocity with radius.

Finally, it will be interesting to explore the implications of the disk distortions for the shape of the dark halo. Throughout this paper we have only discussed the shape of the total potential. Since loop orbits "oppose" the shape of their potential, our results are consistent with halos that are quite distorted (in the disk plane); this is true in particular if the disks contribute much to the total gravitational force.

**Acknowledgments:** It is a pleasure to acknowledge Penny Sackett, Marijn Franx, Tim de Zeeuw, David Spergel and Simon White for reading drafts of this paper and for helpful comments. We are grateful to Buell Jannuzi for pointing us to the statistical treatment of non-negative quantities. We acknowledge the Astrophysical Research Consortium for its support of the Detector Development Consortium in the production of the NICMOS 3 detector used in this work. We thank the Rockwell scientists for their efforts in developing these excellent HgCdTe arrays and the NSF for support. These data were also in part obtained with a charge-coupled device developed by Tektronix for the Space Telescope Imaging Spectrograph (STIS) program for the Hubble Space Telescope. We also thank the Las Campanas mountain staff for its usual terrific observing support. Financial support for HWR and DZ was provided by NASA through grants HF-1024.01-91A and HF-1027.01-91A, respectively, from STScI which is operated by AURA, Inc., under NASA contract NAS5-26555.

# References

Athanassoula, L., 1984, Phys. Reports 114, 319

Baldwin, J., Lynden-Bell, D. and Sancisi, R., 1980, MNRAS, 193, 313

Begeman, K. 1987, Ph.D. thesis, Groningen

Binney, J. and de Vaucouleurs, G., 1981, MNRAS, 194, 679

Binney, J. and Tremaine, S. 1987, "Galactic Dynamics", (Princeton University Press: Princeton)

Blitz, L. and Spergel, D. 1990, Ap.J., 370, 205

Charlot, S. and Bruzual,' A.G., 1991, Ap.J., 367, 126

Contopoulos, G. and Grosbøl, P. 1988, AA, 197, 83

de Vaucouleurs, G. et al. 1991, "Third Reference Catalogue of Bright Galaxies," (Springer-Verlag: New York).

Elmegreen, D. M., 1981, ApJS, 47, 229

Elmegreen, D. M. and Elmegreen, B. G., 1984, ApJS, 54, 127

Elmegreen, B. G. and Elmegreen, D. M., 1985, ApJ, 288, 438

Elmegreen, B. G., Elmegreen, D. M., and Seiden, P. 1989, Ap. J. 343 602

Elmegreen, B. G., Elmegreen, D. M., and Montenegro, L. 1992, Ap. J. Supp. 79, 37

Elmegreen, B. G., Elmegreen, D. M., and Montenegro, L. 1993, PASP, 105, 644

Franx, M. and de Zeeuw, T. 1992, Ap.J.Lett. 392, L47

Grosbøl, P. 1987, in "Selected Topics on Data Analysis in Astronomy," ed. L. Scarsi, V. DiGesu, and P. Crane (Singapore: World Scientific), p. 57

Grosbøl, P. 1993, PASP, 105, 651

Huchra, J. 1993, CfA Redshift Catalogue

Huizinga, J and van Albada, T., 1992, MNRAS, 254, 677

Huchtmeier, W.K. and Richter, O.-G. 1989, *HI Observations of Galaxies* (New York: Springer)

Kennicutt, R.C. 1983, Ap.J. 272, 54

Kennicutt, R.C. and Edgar, B. K., 1986, Ap.J. 300, 132

Kennicutt, R.C., Tamblyn, P. and Congdon, C. W, 1994, Ap.J. *in press*

Kuijken, K. 1993, Ap.J. 409, 68

Kuijken, K. and Tremaine, S. 1994, Ap.J. 421, 178

Matthewson, D., Ford, V. and Buchhorn, M., 1992, Ap.J.Supp. 81, 413

Lewis, B. M., 1984, Ap.J., 285, 453

Pierce, M.J. and Tully, R.B. 1992, Ap. J. 387, 47

Press, W.H., Flannery, B.P., Teukolsky, S.A., Vatterling, W.T. 1986, *Numerical Recipes*, (Cambridge University: New York)

Regan, M. W. and Wilson C.D. 1993, AJ, 105, 499

Rix, H.-W. and Rieke, M.J. 1993, Ap.J., 418, 123

Rix, H.-W. and White, S.D.M. 1993

Sandage, A. and Tammann, G.A. 1981, *A Revised Shapley-Ames Catalogue of Bright Galaxies*, (Washington: Carnegie Institute of Washington).

Sandage, A. 1961, *The Hubble Atlas of Galaxies*, (Washington: Carnegie Institute of Washington).

Schommer, R., Bothun, G., Williams, T. and Mould, J., 1993, A.J. 105, 97

Schweizer, F. 1976, Ap. J. Supp. 31, 313

Jensen,, E. B., Talbot, R. J. and Dufour, R. J., 1981, Ap.J., 243, 716

Tully, R.B. 1988, *Nearby Galaxies Catalog*, Cambridge University: Cambridge

Tully, R. B. and Fisher, J. R. 1977, A.A., 54, 661

Visser, H. C. D. 1980, A&A, 88, 149

Wardle, J. F. C. and Kronenberg, P. P. 1974, Ap.J., 194, 249

Warren, M., Quinn, P., Salmon, J. and Zurek, W. 1992, Ap.J. 339, 405

Appendix A. Model Fitting

Here we describe in detail the parameterized model for the light distribution in the observed disk galaxies and the procedure used to find the best-fit parameters. This model is used in Section 3.2 to determine the disk ellipticities.

A1. The Bulge-Disk Model

Our model for the light distribution of a sample galaxy consists of two components, a bulge and a disk, each of which is characterized by a parameterized fitting function. The bulge is assumed to have an $R^{1/4}$ luminosity profile and constant ellipticity. The disk is assumed have an exponential luminosity profile and an ellipticity profile as expected for a constant potential ellipticity.

The components are modeled with simple analytic expressions. The intensity at any point $(R, \phi)$ is given by

$$I(R, \phi) = I_B(R, \phi, \vec{p}) + I_D(R, \phi, \vec{p}) + I_{sky}, \quad (A1)$$

where $\vec{p}$ denotes the vector of the model parameters, the origin of the coordinate system is assumed to be at the galaxy center, and $I_B$ and $I_D$ are the bulge and disk models respectively. The bulge model is given by

$$I_B(R, \phi) = I_e \cdot \exp\left\{ -7.67 \left[ \left( \frac{R_B(R, \phi)}{R_e} \right)^{1/4} - 1 \right] \right\}, \quad (A2)$$

where

$$R_B(R, \phi) = R \cdot \sqrt{(1 - \epsilon_B) \sin^2(\phi) + \frac{1}{1 - \epsilon_B} \cos^2(\phi)}, \quad (A3)$$

$I_e$ is the effective intensity, $R_e$ is the effective radius, $\epsilon_B$ is the ellipticity, and $\phi_B$ is the minor axis position angle. The disk model is given by

$$I_D(R, \phi) = I_0 \exp\left\{ -\frac{R_D(R, \phi)}{R_{exp}} \right\}, \quad (A4)$$

where $I_0$ is the central disk surface brightness, $R_{exp}$ is the exponential scale radius, $R_D$ is given by

$$R_D(R, \phi) = R \cdot \left( 1 + \tilde{\epsilon}_{iso}(R) \cdot \cos[2(\phi - \phi_D)] \right) \quad (A5)$$

and

$$\tilde{\epsilon}_{iso}(R) = \epsilon_\Phi \cdot \left( 1 + \frac{R}{R_{exp}} \right). \quad (A6)$$

The decrease of the apparent eccentricity with increasing radius is expected for a disk embedded in a potential with constant distortion (Franx and de Zeeuw 1992). The sky level, $I_{sky}$, should be retained as a free variable, because the galaxies have a radial extent comparable to the field covered by the detector.

A2. Determining the Best Parameters

A2.1. Rebinning the Image

The "best fit model" is defined as the set of parameters, $\vec{p}_{best}$, which minimizes

$$\chi^2(\vec{p}) \equiv \sum_{all\ data\ points} \left( \frac{I(R,\phi,\vec{p}) - I_{data}(R,\phi)}{\sigma_{data}(R,\phi)} \right)^2. \quad (A8)$$

With the algorithm employed here to find $\vec{p}_{best}$, this sum must be evaluated many times ($\sim 1000$). If this sum is taken over all pixels of a $\sim 250 \times 250$ image it contains 62500 terms. A factor of $\sim 100$ in computing time can be saved if the image is rebinned onto a $(\log(R)\text{-}\phi)$ grid. Little information is lost in such rebinning because images of face-on galaxies have a geometry which is compatible with cylindrical coordinates and because the models have no features that cannot be captured by a $\sim 30 \times 24$ $(\log(R)\text{-}\phi)$ grid. Because pixels with $R < 6$ pixels ($= 2.5''$) are affected by seeing, we restrict the polar grid to $R > 6$ pixels. For $R \gtrsim 10$ pixels the $(\log(R)\text{-}\phi)$ bins are statistically independent and $\chi^2$ tests may be applied. A grid of $30 \times 24$ was found to have sufficient resolution ensure that neither the model nor the data vary significantly within a bin.

A2.2. Biased Random Walk

Because the parameter space spanned by the model has a high dimension and is infinite, the problem of finding the global minimum of $\chi^2$ at $\vec{p}_{best}$ is difficult. Standard techniques of starting at an initial parameter guess, $\vec{p}_{init}$, and using a downhill gradient search (e.g. Press et al. 1986) are unreliable because it is unlikely that from a given point $\vec{p}_{init}$ the path along the local $\chi^2$ gradient will lead to $\vec{p}_{best}$, rather than to a local extremum. Furthermore, we want to find the best solution for the parameters, subject to certain constraints, such as $I_e, I_0, R_e, R_{exp}, \epsilon_B, \epsilon_D \geq 0$. It is difficult to implement such constraints in gradient search methods.

To avoid these difficulties, we use a biased random walk technique (often called simulated annealing, see Press et al. 1986), described in Rix and White (1992), that has proven to be a much more robust, though computationally expensive, means of finding the global minimum of $\chi^2$, independent of the starting parameter guess, $\vec{p}_{init}$. The application of the technique begins by selecting a "box" in parameter space defined by setting $p_i^{min}$ and $p_i^{max}$ for all parameters on the basis of prejudice of permissible parameter ranges or physical constraints. Trial steps, $\delta \vec{p}$, are then taken in parameter space starting from $\vec{p}_{init}$. At each step, $\chi^2$ is evaluated and compared to $\chi^2$ at the previous point in parameter space. The components of the steps, $\delta p_i$, are drawn randomly between $-\delta p_i^{max}$ and $\delta p_i^{max}$. The vector of maximal steps, $\delta \vec{p}^{max}$, must be specified. Such a step is always accepted in the approach to $\chi^2_{min}$ if $\Delta \chi^2 < 0$. However, the step is also accepted with a probability of $\exp(-\Delta \chi^2 / \Delta \chi_0^2)$ if $\Delta \chi^2 > 0$, where $\Delta \chi_0^2$ is an adjustable "uphill penalty." All steps that would cause an exit of the parameter box are rejected. For a large initial value of $\chi^2$ a few hundred to a few thousand steps are necessary. Once in the neighborhood of $\chi^2_{min}$, the $\chi^2$ penalty is stiffened and the maximal step sizes are decreased (see also Rix and White 1992). Finally, $\chi^2_{min}$ is approached with a conventional gradient technique (Press et al. 1986) which enables us to calculate the co-variance matrix.

Appendix B. The Age Distribution of Stars Observed in the B and K Bands

Here we estimate the contribution of "young" stars to the total light of the galaxy at B and K($2.2\mu$m) with a simple model. In this context, "young" means that $t_{age}$, the age of the stars, satisfies $t_{age} \lesssim$ few $t_{dynamical} \sim 3 \times 10^8$yrs. The K-band results are of immediate relevance to this paper, the B-band results are given for comparison. We use $\langle L_0 \rangle_\lambda$ to denote the mean luminosity at birth of a set of stars for a given IMF, and $s(t)$ to denote the rate at which stars, now at age $t$, were formed. At any given wavelength $\lambda$, we define the fading of this set of stars with age as $f_\lambda(t)$ such that

$$L(t)_\lambda = f_\lambda(t)\langle L_0\rangle_\lambda, \qquad (C1)$$

where $L(t)_\lambda$ is the luminosity at age $t$. The total luminosity due to stars younger than $t$ can then be written as

$$L_\lambda(<t) = \langle L_0\rangle_\lambda \int_0^t s(\tau)f_\lambda(\tau)d\tau, \qquad (C2)$$

and the fraction of the luminosity they contribute to the total luminosity is

$$R_\lambda(<t) = L_\lambda(<t)/L_\lambda(<t_{age}), \qquad (C3)$$

where for a current galaxy $t_{age} = t_0 \sim 10^{10}$ years.

To evaluate this integral exactly we need to know the star formation history and the fading function, which could be derived from stellar evolution models. To obtain a rough estimate of $L_\lambda(<t)$, we make two simplifications: (1) we set $s(t) = s_0 + (2s_0/t_{age})t$ (Kennicutt 1983, Kennicutt et al.1994); and (2) we set

$$f(t) = \begin{cases} f_0, & \text{for } t < 10^7 \text{yrs}; \\ (t/10^7)^{-\gamma}, & \text{for } t > 10^7 \text{yrs}, \end{cases}$$

where for the B-band $f_0 \sim 1$ and $\gamma \sim 0.84$, for the K-band $f_0 \sim 3$ (to allow for the so-called red excursion of supergiant stars in the HR diagram), and $\gamma \sim 0.57$. The latter quantities were extracted from the Figures given in Charlot and Bruzual (1991). We now express the luminosity of stars younger than $t$ as

$$L_\lambda(<t) = \langle L_0\rangle_\lambda\ s_0 \left[f_0 + \frac{t^{1-\gamma}-1}{1-\gamma} + \frac{2(t^{2-\gamma}-1)}{(2-\gamma)t_{age}}\right]. \qquad (C4)$$

From this, $R_\lambda(<t)$ is easily calculated and is plotted in Figure 6 for both the B and K bands.

## Appendix C. Non-parametric Estimate of the Ellipticity Distribution

Here we discuss how to estimate the most likely distribution of intrinsic disk ellipticities, $P(\epsilon_\Phi^I)$, given a set of measured ellipticities, $\epsilon_\Phi$, with error $\Delta\epsilon$, for a galaxy observed at a kinematic inclination $\cos(i)$. The second angle, $\phi$, determining the disk orientation at fixed inclination $i$, is not known. Hence we must assume it to be distributed randomly. Let $\tilde{\epsilon}_j$ be the mean intrinsic ellipticity of the isophotes of the $j^{th}$ galaxy disk. Since most of the statistical weight in our fits of $\epsilon_{\Phi,j}$ arises from radii $R_{exp} < R < 3R_{exp}$, we assume that $\tilde{\epsilon}_j = 1.5\epsilon_{\Phi,j}$ (see Equation 11). An isophote can then be labeled by the parameter $\psi$ as

$$\vec{r} = \cos(\psi)\hat{x} + (1-\tilde{\epsilon}_j)\sin(\psi)\hat{y} \qquad (B1)$$

and the viewing direction can be written as

$$\hat{n} = \cos(\phi)\sin(i)\hat{x} + \sin(\phi)\sin(i)\hat{y} + \cos(i)\hat{z}. \qquad (B2)$$

Hence, at any point $\psi$ along the streamline the projected radius vector is given by $\vec{r}_{proj} = \vec{r}\times\hat{n}$ with length

$$r_{proj}^2 = \cos^2(i)\left(\cos^2(\psi)+(1-\tilde{\epsilon}_j)^2\sin^2(\psi)\right) + \sin^2(i)\left(\cos(\psi)\sin(\phi)-(1-\tilde{\epsilon}_j)\sin(\psi)\cos(\phi)\right)^2. \qquad (B3)$$

The ellipticity of the ellipse formed by the streamline is

$$\tilde{\epsilon}_j = 1 - r_{proj}(min)/r_{proj}(max). \qquad (B4)$$

We expect to measure (even without noise) $\tilde{\epsilon}_{\Phi,j} = \tilde{\epsilon}_j/1.5$. In the presence of a finite measurement error $\Delta \epsilon_j$, the distribution, $\tilde{P}(\epsilon_{\Phi,j})$, of expected measurements is given by

$$\tilde{P}(\epsilon_{\Phi,j}, \tilde{\epsilon}_{\Phi,j}, \Delta\epsilon_j) = \frac{\epsilon_{\Phi,j}}{\Delta\epsilon_j^2} I_0\left(\frac{\epsilon_{\Phi,j}\tilde{\epsilon}_{\Phi,j}}{\Delta\epsilon_j^2}\right) \exp\left(-\frac{\epsilon_{\Phi,j}^2 + \tilde{\epsilon}_{\Phi,j}^2}{2\Delta\epsilon_j^2}\right). \qquad (B5)$$

This distribution is called the Rice distribution (see Wardle and Kronenberg 1974). For $\tilde{\epsilon}_{\Phi,j} \gg \Delta\epsilon_j$ it tends to a Gaussian about $\tilde{\epsilon}_{\Phi,j}$ with dispersion $\Delta\epsilon_j$. When the underlying ellipticity is small compared to the measurement error, $\tilde{\epsilon}_{\Phi,j} \ll \Delta\epsilon_j$, the expectation value for a measurement does not tend to zero but rather to $\Delta\epsilon_j$. Note that the expression in B5 depends on the angles $\phi$ and $i$ through $\tilde{\epsilon}_{\Phi,j}(\phi, i)$. Averaging this distribution (B5) over all angles $\phi$ yields the probability distribution, $P_j(\epsilon_\Phi^I)$, for the intrinsic ellipticity for the $j^{th}$ object. The final probability distribution, $P(\epsilon_\Phi^I)$, is obtained by averaging $P_j(\epsilon_\Phi^I)$ over our sample of galaxies, and this distribution is shown in Figure 7. Confidence regions for $P_j(\epsilon_\Phi^I)$ were estimated by using the bootstrapping technique (Press et al.1993). The technique consists of drawing 18 random members from the sample, allowing for multiple draws of individual sample members, and repeating the above analysis. By repeating many times this allows one to build an ensemble of possible $p(\epsilon_{intr})$ distributions with the same statistics as an ensemble drawn from the true $p(\epsilon_{intr})$. From the distribution of $p(\epsilon_{intr})$, we calculate confidence limits on $p(\epsilon_{intr})$ for each value of $\epsilon_{intr}$. The derived curve and error bounds are shown in Figure 7.

TABLE 1 : The Sample

| Name | Type | $m_B$ | D(′) | $M_B$ | $v_{HI}$[km/s] | $W'_{20}$[km/s] | $i(°)$ | $t_e(s)$ |
|---|---|---|---|---|---|---|---|---|
| NGC 600 | SBd | 12.9 | 3.3 | −19.0 | 1843 | 80 | 17 | 1800 |
| NGC 991 | Sc | 12.4 | 2.7 | −19.2 | 1535 | 82 | 17 | 1800 |
| NGC1015[a] | SBa | 13.0 | 2.6 | −19.7 | 2630 | 87 | 15 | 1800 |
| NGC1302[b] | Sa | 11.6 | 4.8 | −20.2 | 1703 | 105 | 16 | 1620 |
| NGC1309[b] | Sbc | 12.0 | 2.2 | −20.3 | 2135 | 156 | 23 | 1800 |
| NGC1325A | Sb | 13.3 | 2.7 | −17.9 | 1333 | 46 | 14 | 1800 |
| NGC1376[b] | Scd | 12.8 | 2.0 | −21.0 | 4162 | 179 | 21 | 1800 |
| NGC1642[b] | Sc | 13.3 | 2.1 | −20.6 | 4621 | 143 | 19 | 1800 |
| NGC1703[b] | SBb | 11.9 | 4.5 | −19.6 | 1526 | 65 | 11 | 1980 |
| NGC2466[b] | Sc | 13.5 | 1.8 | −20.7 | 5364 | 175 | 23 | 1800 |
| NGC2485[c] | Sa | 13.1 | 1.6 | −20.8 | 4612 | 200 | 25 | 1620 |
| NGC2718 | Sab | 12.7 | 2.1 | −20.8 | 3843 | 130 | 16 | 1800 |
| NGC6814 | Sbc | 12.1 | 3.0 | −19.5 | 1565 | 112 | 21 | 1080 |
| NGC7156 | Scd | 13.1 | 1.6 | −20.5 | 3984 | 125 | 17 | 1500 |
| NGC7309 | Sbc | 13.0 | 1.9 | −20.6 | 4000 | 136 | 18 | 1800 |
| NGC7742 | S0 | 12.4 | 2.0 | −19.4 | 1649 | 85 | 17 | 2160 |
| IC 2627 | Sbc | 12.6 | 4.0 | −19.6 | 2082 | 42 | 8 | 360 |
| ESO-436 | Sc | 13.4 | 1.8 | −20.2 | 4079 | 85 | 13 | 1260 |

[a] HI profile remeasured by Tully (see Huchtmeier and Richter, 1989).
[b] I-band image was obtained.
[c] HI profile has broad wings.

TABLE 2 : Results from Fourier and Model Fitting Analysis

| Name[a] | $R_{exp}$(arcsec) | $\tilde{A}_1(2.5R_{exp})$ | $\epsilon_\Phi$ | PA(°) |
|---|---|---|---|---|
| NGC 600 | 17.6 ± 2.8 | 0.11 ± 0.08 | 0.09 ± 0.06 | 177 ± 18 |
| NGC 991 | 19.5 ± 2.0 | 0.09 ± 0.05 | 0.10 ± 0.03 | 89 ± 09 |
| NGC1015 | 15.2 ± 1.5 | 0.23 ± 0.08 | 0.04 ± 0.03 | 116 ± 08 |
| NGC1302 | 11.2 ± 0.8 | 0.03 ± 0.02 | 0.18 ± 0.03 | 171 ± 03 |
| NGC1309 | 11.1 ± 0.6 | 0.37 ± 0.07 | 0.09 ± 0.01 | 11 ± 03 |
| NGC1325A | 21.5 ± 3.0 | 0.58 ± 0.25 | 0.06 ± 0.07 | 163 ± 55 |
| NGC1376 | 11.4 ± 0.6 | 0.05 ± 0.08 | 0.04 ± 0.03 | 152 ± 23 |
| NGC1642 | 9.4 ± 0.4 | 0.01 ± 0.02 | 0.04 ± 0.01 | 117 ± 08 |
| NGC1703 | 18.0 ± 1.5 | 0.17 ± 0.08 | 0.06 ± 0.04 | 147 ± 16 |
| NGC2466 | 7.7 ± 0.3 | 0.17 ± 0.04 | 0.10 ± 0.02 | 7 ± 02 |
| NGC2485 | 14.5 ± 2.4 | 0.27 ± 0.10 | 0.11 ± 0.05 | 174 ± 10 |
| NGC2718 | 14.1 ± 3.3 | 0.00 ± 0.07 | 0.27 ± 0.07 | 156 ± 06 |
| NGC6814 | 18.6 ± 0.6 | 0.19 ± 0.04 | 0.09 ± 0.02 | 9 ± 05 |
| NGC7156 | 9.2 ± 0.4 | 0.08 ± 0.04 | 0.06 ± 0.02 | 40 ± 11 |
| NGC7309 | 12.7 ± 0.4 | 0.19 ± 0.05 | 0.12 ± 0.02 | 153 ± 04 |
| NGC7742 | 6.6 ± 0.2 | 0.05 ± 0.01 | 0.02 ± 0.01 | 162 ± 15 |
| IC 2627 | 21.2 ± 1.2 | 0.17 ± 0.04 | 0.07 ± 0.03 | 28 ± 10 |
| ESO-436 | 14.8 ± 1.8 | 0.16 ± 0.14 | 0.10 ± 0.07 | 164 ± 20 |

Figure Captions

Figure 1: K$'$ images of the 18 galaxies in the sample. The angular size of each frame is $1.73' \times 1.73'$. The galaxies are shown in the same order as listed in Table 1 from left to right and top to bottom. North is at the top and East is to the left in all frames. Except for NGC 1302 and NGC 6814 all galaxies are displayed with the same stretch.

Figure 2: Amplitudes and phases of the principal Fourier components in the K$'$ images of the sample galaxies. Radius is plotted in disk scale-length units. Note that the amplitude scale on the bottom three panels has been expanded by 5/3.

Figure 3: Histogram of disk lopsidedness for galaxies in the sample. The Fourier power $\tilde{A}_1$ measured at 2.5 exponential disk scale lengths is expressed along the abscissa. The values of $\tilde{A}_1$ have been corrected for their measurement errors (see text).

Figure 4: Comparison between I and K$'$ results for the six galaxies for which I-band data was obtained. The K$'$ $-$I color, and the phases and amplitudes of the $m = 1$ and 2 components are shown (solid circles represent I data, open circles represent K$'$ data).

Figure 5: Model galaxy image, with $m = 2$ and 4 spirals characteristic of NGC 1703. The axisymmetric input disk and the arms result in an estimated disk ellipticity of only 0.03.

Figure 6: The age distribution of the stellar K$'(2.2\mu$m)-flux (open symbols) and B -flux (filled symbols), as predicted by the models described in Appendix C based on Charlot and Bruzual (1991). The top panel shows the predicted fraction of the total flux at current time ($t_{age} = 10^{10}$ yrs) arising from stars younger than $t$. Note that even in K$'(2.2\mu$m) the fraction of light from stars younger than a few dynamical periods, $t \sim 3 \times 10^8$ yrs, may be 10%. The bottom panel shows the fraction of the luminosity from stars younger than $t$ that arises from stars younger than $0.5t$.

Figure 7: The distribution of potential ellipticities $P(\epsilon_\Phi^I)$. The solid line delineates the $P(\epsilon_\Phi^I)$ distribution calculated from the data and the long dashed lines delineate the region of 95% confidence as estimated by "bootstrapping" (cf. Appendix B). The dotted line delineates the expected distribution from axisymmetric model disks at the inclinations given in Table 1 and measured with the observational errors estimated for our sample.

Figure 8: The histogram of measured ellipticities, $\epsilon_\Phi$, and the expected distribution of such measurements for three assumed *intrinsic* ellipticity distributions are shown. The solid line assumes that $P(\epsilon_\Phi^I, \epsilon_0) = 1/\epsilon_0 \, \exp(-\epsilon_\Phi^I/\epsilon_0)$ with $\epsilon_0 = 0.045$. The more highly peaked dashed line illustrates the result from the same functional form but with $\epsilon_0 = 0.025$, the other dashed line the result from a model with $\epsilon_0 = 0.075$.

Figure 9: Azimuthal intensity variations, as characterized by $(\Sigma_{max} - \Sigma_{min})/\Sigma_{min}$. In order to focus on variations with two-fold symmetries, the maximal and minimal surface brightness were not calculated from the image itself, but from its Fourier components with $m = 0, 2, 4, 6$. The majority of galaxies show azimuthal variations in excess of unity at least at some radii. In some objects, e.g. NGC 1015, such strong variations are due to a bar, for many others, e.g. NGC 1309, NGC 2485, NGC 6814, NGC 7309, they are predominantly due to two-arm spirals.

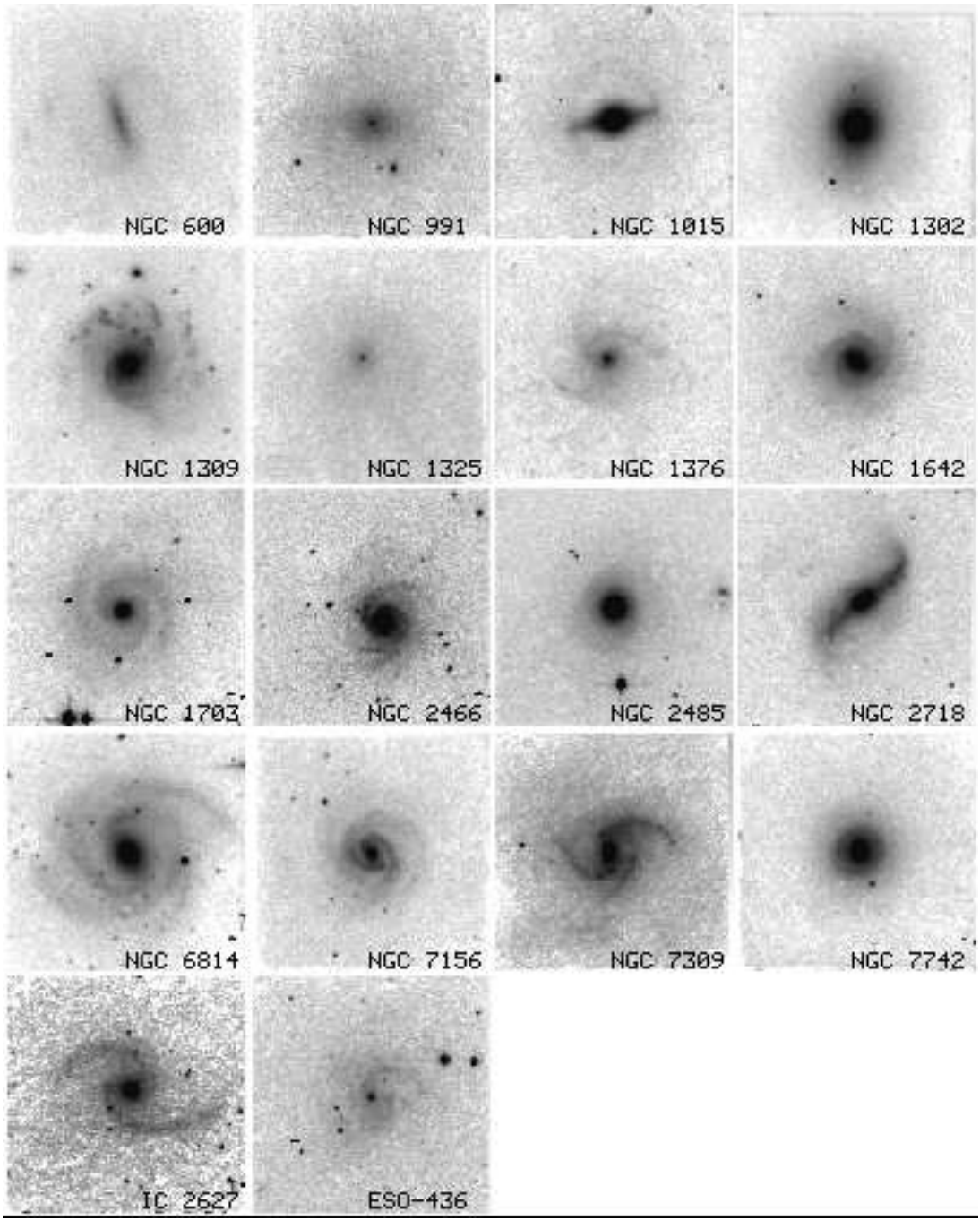

28